# β-Hydroxybutyrate remodels the C99 interactome and coincides with restored organelle homeostasis in a *Drosophila* Alzheimer's model.

Hao Huang[a], Kaijing Xu[a] and Michael Lardelli[a]

## Abstract

Early endolysosomal and autophagic defects are among the earliest cellular alterations observed in Alzheimer's disease (AD), yet the molecular drivers linking amyloid precursor protein (APP) metabolism to vesicle trafficking dysfunction remain incompletely understood. The APP-derived fragment C99 has emerged as a potential upstream mediator of intracellular toxicity, but its impact on organelle homeostasis and its modulation by metabolic interventions remain unclear. Here, we show that neuronal expression of human C99 in *Drosophila* induces profound vesicular abnormalities, impaired autophagic turnover, and disrupted mitochondrial quality control. Ultrastructural analysis revealed extensive accumulation of enlarged vesicular compartments, accompanied by reduced mitochondrial turnover and accumulation of aged mitochondria. Treatment with the ketone body β-hydroxybutyrate (BHB) restored autophagic cargo clearance, improved mitochondrial turnover, and normalized vesicular ultrastructure. These protective effects required neuronal ketone transport, indicating a neuron-intrinsic metabolic mechanism. Proteomic mapping of the C99-associated interactome revealed that ketone treatment remodels networks enriched for vesicle trafficking and proteostasis pathways. Network prioritization identified the retromer component VPS35 as a candidate regulatory hub. Functional analyses demonstrated that depletion of VPS35 abolished the BHB-dependent restoration of autophagy, mitochondrial turnover, and vesicle morphology. Together, these findings suggest that ketone treatment restores mitochondrial quality control through a VPS35-dependent mechanism in C99-induced neurodegeneration, providing mechanistic insight into how metabolic interventions may restore intracellular homeostasis in Alzheimer's disease.

Keywords: Alzheimer's disease; β-hydroxybutyrate; C99; Mitophagy; Metabolism; Drosophila

Author affiliations:

a Faculty of Sciences, Engineering and Technology, University of Adelaide, North Terrace, Adelaide, 5005, SA, Australia.

Correspondence to: Hao Huang

Full address：Gate8 Victoria Dr, The University of Adelaide Adelaide, North Terrace, SA5005, Australia

E-mail：hao.huang01@adelaide.edu.au

**Running title**: Remodeling of C99-associated protein interactions by BHB

**Keywords:** Alzheimer's disease; $\beta$-Hydroxybutyrate (BHB); C99; Proteostasis

# Introduction

Alzheimer's disease (AD) is the most common cause of dementia and represents a major global health challenge. It is characterized by progressive cognitive decline accompanied by synaptic dysfunction and neuronal loss[1,2]. Pathologically, AD is associated with extracellular amyloid plaques and intracellular neurofibrillary tangles[3]. Despite decades of research, therapies targeting these classical pathological features have shown limited success, highlighting the need to better understand the cellular mechanisms that drive neuronal dysfunction in AD.

The amyloid precursor protein (APP) undergoes sequential proteolytic processing that generates several fragments, including the β-secretase–derived C-terminal fragment C99[4]. Although amyloid-β (Aβ) has long been considered the primary pathogenic species in AD, increasing evidence indicates that C99 itself can exert toxic effects independently of Aβ production. Elevated C99 levels have been associated with alterations in endosomal trafficking, cholesterol homeostasis, and lysosomal function—processes closely linked to endolysosomal and autophagic pathways[5–7].

Autophagy plays a central role in maintaining neuronal proteostasis and organelle quality control. Efficient autophagic flux requires coordinated formation of autophagosomes, their maturation, and subsequent fusion with lysosomes for cargo degradation[8]. Neurons in AD frequently exhibit accumulation of autophagic vesicles, indicating impaired autophagic flux rather than simple activation of autophagy[9–11]. Disruption of endolysosomal trafficking and lysosomal function has been proposed as a major contributor to this defect[12,13]. Among autophagy-dependent quality control pathways, mitophagy is particularly important for neuronal health, as damaged mitochondria must be selectively removed to prevent oxidative stress and metabolic imbalance[7]. Emerging evidence suggests that APP-derived fragments, including C99, may interfere with endolysosomal and autophagic pathways, potentially leading to defective autophagosome maturation or impaired mitophagy. However, the molecular mechanisms through which C99 disrupts autophagic flux and mitochondrial quality control remain poorly understood.

In parallel with efforts to understand pathogenic mechanisms, metabolic interventions have attracted increasing interest in neurodegenerative disease research. Ketone bodies, including β-hydroxybutyrate (BHB), serve as alternative metabolic substrates during conditions such as fasting or ketogenic diets[14]. In addition to their role in energy metabolism, ketone bodies can function as signaling molecules that regulate pathways involved in cellular stress responses and mitochondrial function[15–17]. BHB has been reported to activate AMPK signaling, inhibit mTOR activity, and modulate histone deacetylase activity, pathways closely linked to the regulation of autophagy and mitophagy[18,19]. Ketogenic interventions and ketone supplementation have been associated with improved neuronal resilience and cognitive performance in experimental models of AD[20]. However, the molecular targets through which ketone metabolism influences disease-relevant cellular pathways remain incompletely defined.

These observations raise an important question: can metabolic signals such as ketone bodies modulate the cellular defects induced by C99 accumulation? In particular, it remains unclear whether ketone metabolism can restore mitophagy and autophagic flux disrupted by C99.

Here, we addressed this question using a Drosophila model expressing C99. We examined how C99 accumulation affects mitochondrial quality control and autophagic flux in vivo and tested whether ketone treatment can alleviate these defects. To explore the underlying mechanisms, we analyzed the C99-associated protein interaction network using proteomic approaches and integrative computational analysis. Together, these studies identify candidate regulatory pathways linking C99 signaling to autophagy and mitochondrial quality control, and provide insight into how metabolic state may influence cellular pathology in Alzheimer's disease.

# Materials and methods

## *Drosophila* Stocks

The following *Drosophila* stocks were used in this study, maintained under a 12-hour light/12-hour dark cycle at 60% relative humidity: GMR-Gal4 (BDSC #1104), UAS-C99 (BDSC #33783; BDSC #33784), UAS-LC3-GFP (BDSC #8730), UAS-mito-QC (BDSC #91641), VPS35-GD (v22180, v45570), sln-GD (v4607), chk-GD (v37139), elav2-mito-Timer (fluorescent group consistent with BDSC #57323) and UAS-GFP-Ref(2)P (consistent with BDSC # 605356). These stocks were either generated, maintained, or provided by Ms. Louise O'Keefe for this study. To ensure genetic background consistency, all UAS and GAL4 lines were backcrossed more than 5 generations into $w^{1118}$ prior to crossing. Where backcrossing was not applicable (e.g., BDSC standard lines), parental controls were maintained on the same $w^{1118}$ background.

## *Drosophila* Treatment

For this study, AD *Drosophila* models were generated by expressing human C99 under the vitreous GMR-Gal4 driver. Flies were maintained at 18°C until eclosion, after which adults were transferred to 25°C to initiate gene expression and treatment. BHB (2 mM, Sigma Aldrich, Cat. #: H6501) treatments were administered by transferring larvae to supplemented food treat at least 10 days after emergence. The BHB dosage was derived from previous experiments[21]. For any results presented, the fruit flies were female fruit flies 13 days after emergence.

## Transmission Electron Microscopy

Adult *Drosophila* were anesthetized in cold phosphate-buffered saline (PBS) and dissected under a stereomicroscope. Photoreceptor tissues from the compound eye were isolated for ultrastructural analysis due to their highly ordered cellular architecture, which enables consistent anatomical sampling across individuals. Dissected head tissues were immediately fixed in a solution containing 4% paraformaldehyde, 1.25% glutaraldehyde (EM grade), and 4% sucrose (pH 7.2) at 4 °C overnight. Samples were washed twice in PBS supplemented with 4% sucrose for 10 min each and post-fixed in 2% osmium tetroxide ($OsO_4$) for 45 min at room temperature to enhance membrane contrast.

Samples were dehydrated through an ethanol series (70%, 95%, and 100%), with each step repeated three to four times for 10 min. Resin infiltration was performed by incubating tissues in a 1:1 mixture of propylene oxide and resin for 1 h at room temperature, followed by immersion in 100% resin overnight. Polymerization was carried out at 70 °C for 48 h. Ultrathin sections (~70 nm) were prepared using an ultramicrotome and mounted on copper grids. Sections were stained with 2% uranyl acetate followed by 0.5% lead citrate to enhance contrast. Imaging was performed using a FEI Glacios 200 kV cryo-transmission electron microscope at 4800× magnification, focusing on photoreceptor cells within the Drosophila retina.

Mitochondrial segmentation was performed using **ilastik** (v1.4.0). A publicly available pre-annotated dataset of electron microscopy images containing dense annotations of neuronal membranes, synapses, mitochondria, and extracellular space was used to train the segmentation model. Within the Pixel Classification workflow, annotated mitochondrial regions served as ground truth for supervised training of a Random Forest classifier. The classifier was trained using intensity, edge, and multi-scale texture features to distinguish mitochondria from surrounding cellular structures. The trained model was then applied to the entire dataset to generate probability maps, which were thresholded to produce binary segmentation masks.

Post-processing and morphological analysis were performed using OpenCV. Segmented mitochondrial contours were detected, and morphological parameters including size, shape, and spatial distribution within photoreceptor cells were quantified. To evaluate vesicular abnormalities, vesicle-like structures were segmented using an area-based thresholding approach. Vesicles with size distributions consistent with lysosomal compartments were selected for further analysis. Final segmentation outputs, including mitochondria and vesicle masks, were overlaid onto the original TEM images for visualization and validation.

## Imaging of *Drosophila* Eyes

Adult *Drosophila* were anesthetized on ice for 10 min or exposed to $CO_2$ for approximately 15 min prior to imaging. Flies were mounted on glass slides using double-sided tape with the head oriented upward to expose the compound eyes. Imaging was performed under low-light conditions to minimize photobleaching and premature fluorescence excitation.

Eye images were first standardized by cropping the compound eye region and normalizing image intensity and resolution. The eye area was segmented using an intensity-based mask to define the region of interest for subsequent quantitative analysis.

To quantify retinal degeneration in Drosophila eyes, we developed a convolutional neural network (CNN)–based image analysis pipeline to estimate eye surface roughness from brightfield images. Eye images were first standardized by cropping the compound eye region and normalizing image intensity and resolution.

The CNN model was trained using a curated dataset of eye images collected from Drosophila experiments conducted at the University of Adelaide Drosophila facility between 2020 and 2024, encompassing a range of normal and degenerated eye phenotypes used for supervised training. The network architecture consisted of convolutional layers for feature extraction followed by fully connected layers for phenotype classification.

To interpret model predictions, Gradient-weighted Class Activation Mapping (Grad-CAM) was applied to generate saliency maps highlighting image regions contributing most strongly to the classification output, typically corresponding to disruptions in the ommatidial lattice.

A quantitative roughness score was calculated from the predicted degeneration probability weighted by the mean Grad-CAM activation within the eye region. Image preprocessing, model training, and analysis were implemented in Python using the PyTorch deep learning framework.

## Fluorescent Imaging of *Drosophila* Eyes

Fluorescence imaging was carried out using a LEICA fluorescence microscope equipped with filter sets optimized for GFP (excitation 488 nm, emission 510 nm) and mCherry (excitation 587 nm, emission 610 nm). GFP and mCherry signals were acquired sequentially to avoid spectral overlap. For longitudinal imaging experiments, flies were re-anesthetized as necessary and returned to fresh food medium following imaging.

## Fluorescence Image Analysis

Fluorescent signals were segmented using adaptive K-means clustering, followed by morphological closing to refine segmented regions. To avoid interference from pseudopupil structures, circular regions corresponding to pseudopupils were automatically excluded using a shape-based mask defined by circularity and intensity thresholds (radius < 15 μm; mean pixel intensity > 1.5× background).

For Mito-QC and MitoTimer analyses, background-corrected fluorescence ratios were calculated as:

$$r = \frac{F_{target} - F_{background}}{F_{reference}}$$

where $F_{\mathrm{target}}$ and $F_{\mathrm{reference}}$ represent the red and green fluorescence intensities, respectively, and $F_{\mathrm{background}}$ is the mean intensity of non-cellular regions. Ratios were normalized to the mean of wild-type controls to allow cross-sample comparison.

## STRING and K-means

Differentially expressed proteins were submitted to the STRING database (v12.0) using gene symbols, specifying *Drosophila* melanogaster as the organism. Protein–protein interaction (PPI) networks were constructed using a minimum interaction confidence score of 0.4. STRING's built-in K-means clustering algorithm was applied to partition the network into functional modules (K = 20). Each cluster was subjected to Gene Ontology (GO), KEGG, and Reactome pathway enrichment analyses, and statistically significant terms (FDR < 0.05) were retained. This automated workflow enabled the identification of coordinated protein groups and their enriched biological functions, as visualized directly within the STRING interface.

## Support Vector Machine

Human gene expression data from the Gene Expression Omnibus (GEO; dataset GSE5281) were used to identify candidate Alzheimer's disease (AD)-related genes using an integrated machine learning framework. Gene expression values were normalized and randomly stratified into training and testing sets.

Feature importance was first estimated using a Random Forest classifier implemented in *scikit-learn*. The top 500 genes ranked by feature importance (importance score > 0.002) were retained and processed using a deep autoencoder neural network (*Keras*) to generate compressed feature representations.

A linear support vector machine (SVM) classifier was then trained on the compressed dataset to estimate gene contributions to AD classification[22]. Model training was repeated 100 times with different random initializations to ensure robustness, and genes with consistently positive SVM weights in ≥80% of resampled datasets were retained.

Human orthologs of Drosophila proteins identified in the C99 interactome were mapped using Ensembl BioMart (release 111) and validated using DIOPT v9.1. Candidate hub proteins were prioritized based on overlap between the machine learning–derived AD gene set and the C99-associated proteomic interactome.

## Immunoprecipitation

This study utilized co-immunoprecipitation (CO-IP) and protein extraction to analyze protein interactions in AD *Drosophila* models. For each genotype or treatment condition, two *Drosophila* heads were homogenized in pre-chilled 1X RIPA buffer supplemented with protease and phosphatase inhibitors. The homogenates were sonicated briefly and incubated on ice for 30 minutes, with vortexing every 10 minutes to enhance lysis. Following incubation, lysates were centrifuged at 14,000 rpm for 10 minutes at 4°C, and the clarified supernatants were collected in pre-chilled microcentrifuge tubes.

CO-IP was performed using the Catch and Release v2.0 Immunoprecipitation Kit (Upstate, Cat. #17-500). Supernatants were incubated overnight at 4°C with an anti-C-terminal antibody (Sigma, Cat. #A8717), following the manufacturer's protocol with modifications to enhance binding efficiency. Specifically, *Drosophila* proteins, antibody, and binding resin were incubated together, allowing for optimized protein capture. Immune complexes were washed three times with pre-chilled wash buffer under gentle rotation at 4°C. Bound complexes were eluted using the provided elution buffer, and eluates were collected for subsequent analysis.

## Mass Spectrometry Analysis

Protein samples were digested into peptides using trypsin (Sigma, Cat. #650279) for subsequent mass spectrometry (MS) analysis. Ice-cold acetone (-20°C) was added at a 4:1 volume ratio, and the mixture was vortexed gently to ensure uniform precipitation. Samples were incubated at -20°C for at least 1 hour or overnight. Following precipitation, proteins were pelleted by centrifugation (12,000–15,000 rpm, 4°C, 10–15 min), and the supernatant was removed. The dried protein pellet was resuspended in 10 mM Tris buffer for MS analysis.

Sample preparation, liquid chromatography-mass spectrometry (LC-MS), and protein identification were performed at the Proteomics Centre, University of Adelaide.

## Plotting and Statistical significance

All statistical analyses were performed using Python (v3.10) with the *SciPy* and *statsmodels* packages. Comparisons among multiple groups were analyzed using one-way or two-way analysis of variance (ANOVA) followed by Tukey's post hoc test for multiple comparisons. Data are presented as mean ± s.e.m., and statistical significance was defined as $p < 0.05$.

All experiments were independently repeated three times with consistent results. Representative data from one experiment are shown in the main figures. n = number of biological replicates for one time.

# Results

## C99 expression induces vesicular accumulation and retinal degeneration

To investigate the cellular consequences of C99 accumulation in vivo, we expressed human C99 in the Drosophila eye using the GMR-GAL4 driver. C99-expressing flies developed a pronounced rough eye phenotype compared with control animals, consistent with retinal degeneration (Fig. 1a–c, Supplementary Fig. S1a).

To examine the underlying cellular alterations, we analyzed retinal ultrastructure using transmission electron microscopy (TEM). Compared to typical two-membrane autophagy, photoreceptors expressing C99 displayed extensive accumulation of enlarged cytoplasmic vesicular structures (Fig. 1d,e, eg. Red box1,2). Many of these vesicles contained densely packed or fibrillar electron-dense material, forming enlarged membrane-bound compartments consistent with alterations in endolysosomal or autophagic pathways [23]. Quantitative analysis confirmed a significant increase in vesicular area and decrease in mitochondria area in C99-expressing retinas compared with controls (Fig. 1g,h).

Interestingly, vesicular structures in BHB-treated C99 retinas exhibited distinct ultrastructural features compared with those observed in untreated C99 retinas. Mitochondrial double-membrane autophagy was rediscovered under the lens(eg. Red box3)[24]. Instead of densely packed luminal material, many vesicles contained multiple small internal vesicles, forming structures morphologically resembling multivesicular endosomal compartments (Fig. 1f, eg. Red box4)[25]. As previous studies have shown, these observations suggest that BHB treatment is associated with alterations in the lysosomal dysfunction[26]. Quantitative results also showed that it rescued the morphological changes in vesicles and mitochondria (Fig. 1g,h).

## C99 expression impairs autophagy and mitochondrial quality control

Given the vesicular abnormalities observed by TEM, we next examined whether autophagic turnover was altered in C99-expressing retinas. Autophagic cargo dynamics were assessed using the GFP–Ref(2)P reporter. Ref(2)P signal was markedly increased in C99-expressing eyes compared with controls, consistent with the accumulation of autophagic cargo (Fig. 2a, b).

To further examine autophagosomes, we analyzed the LC3/Atg8a reporter, which labels autophagosomal membranes. LC3/Atg8a reporter levels were significantly elevated in C99 retinas relative to controls (Fig. 2c, d), consistent with altered autophagosome abundance.

Because mitochondrial turnover is closely linked to autophagy, we next assessed mitophagy using the mito-QC reporter[27,28]. C99 expression led to a significant reduction in the mito-QC red/green ratio, suggesting reduced mitophagy activity (Fig. 2e, f). In agreement with this observation, mitochondrial turnover assessed using the MitoTimer reporter was markedly altered in C99-expressing retinas, consistent with increased mitochondrial age (Fig. 2g, h)[29,30].

## Ketone treatment restores autophagy and mitochondrial turnover in the C99 model

We next asked whether BHB treatment could modify the autophagy- and mitochondria-related alterations observed in C99-expressing retinas. Ref(2)P accumulation was significantly reduced in BHB-treated C99 retinas compared with untreated C99 retinas (Fig. 2a, b), suggesting improved autophagic cargo turnover.

In contrast, LC3/Atg8a reporter levels remained elevated in BHB-treated C99 retinas and were not markedly reduced compared with untreated C99 retinas (Fig. 2c, d), suggesting that BHB treatment does not substantially alter autophagosome abundance. The differential response of Ref(2)P and Atg8a reporters is consistent with improved cargo clearance rather than reduced autophagosome formation.

Importantly, mitophagy activity measured by the mito-QC reporter was significantly increased following BHB treatment (Fig. 2e, f), consistent with increased mitophagy activity. In agreement with this observation, the altered MitoTimer profile observed in C99-expressing retinas was partially normalized after BHB treatment (Fig. 2g, h), suggesting improved mitochondrial turnover.

## Proteomic analysis identifies C99-associated interaction networks and candidate hub proteins

To identify molecular pathways associated with C99 expression, we performed co-immunoprecipitation followed by mass spectrometry (Co-IP/MS) using Drosophila expressing human C99. After subtraction of proteins detected in control samples lacking C99 expression, a set of C99-associated proteins was identified, representing candidate components of C99-containing complexes (Fig. 3a). The background subtraction consisted mainly of keratin derived from highly enriched nucleoproteins and Drosophila chitinous tissue (Fig. 3b). Sample quality and immunoprecipitation efficiency were verified prior to mass spectrometry by immunoblotting and Coomassie staining (Supplementary Fig. S1b, c).

To determine whether ketone treatment alters the C99-associated protein network, we compared interactomes detected in C99 and C99+BHB samples. Changes in the interactome were then evaluated by comparing the presence and relative enrichment of proteins between conditions. Proteins consistently detected in one condition or showing reproducible enrichment across replicates were considered candidate components of a BHB-responsive interactome. These analyses suggest that ketone treatment reshapes the C99-associated protein interaction network.

Functional enrichment and network clustering analyses revealed several major functional modules within the C99-associated protein network, including protein synthesis and RNA processing, metabolic processes, autophagy–proteostasis regulation, and cell cycle–chromosome regulation (Fig. 3c). Notably, the autophagy–proteostasis module contained multiple proteins involved in vesicle trafficking, Rab signaling, and protein quality control pathways. These enriched pathways are consistent with a potential involvement of vesicle trafficking and proteostasis-related processes in C99-associated cellular alterations, which may relate to the vesicular abnormalities and impaired mitochondrial turnover observed in C99-expressing retinas (Figs. 1–2). BHB treatment resulted in most proteins involved in the autophagy pathway no longer being detected, while some proteins related to protein synthesis and RNA function were preserved (Fig. 3d).

Because Alzheimer's disease transcriptomic datasets contain high-dimensional gene expression profiles with substantial redundancy and noise, we employed an integrated machine learning framework to prioritize robust disease-associated features. To prioritize candidate regulatory nodes with potential relevance to Alzheimer's disease, we integrated the C99 interactome with human AD transcriptomic data (GEO: GSE5281). Human orthologs of Drosophila interactors were mapped using Ensembl BioMart and cross-validated using DIOPT[22,31]. This ensemble strategy improves robustness compared with single-method ranking approaches. Gene importance scores derived from a machine learning–based prioritization framework trained on human AD datasets were then used to rank AD-associated genes (Supplementary Fig. S1d). Intersecting these prioritized AD-associated genes with the filtered C99 interactome identified several candidate hub proteins, including VPS35 and PPME1 (Fig. 3e, f).

VPS35, a core component of the retromer complex that regulates endosomal trafficking, was identified within the C99-associated interactome as a candidate hub protein associated with vesicle trafficking and proteostasis pathways[32]. Given the established role of retromer-mediated trafficking in endosomal and autophagy-related processes, these findings nominate

VPS35 as a potential regulatory node linking C99-associated protein networks with vesicle trafficking pathways.

## Neuronal ketone transport mediates BHB-dependent interactome remodeling

To determine whether the effects of BHB reflect a direct ketone-dependent mechanism rather than a general improvement in cellular metabolism, we examined the requirement of ketone transport in the C99 model. In Drosophila, the monocarboxylate transporter Sln mediates neuronal ketone uptake, whereas Chk primarily functions in glial cells[20]. We therefore compared the effects of BHB treatment in flies in which either transporter was selectively inhibited (Fig. 4a).

Proteomic analysis revealed that disruption of neuronal ketone transport by sln knockdown markedly altered the pattern of proteins detected in the C99 interactome under BHB treatment (Fig. 4b). In particular, proteins associated with vesicle trafficking, Rab signaling, and proteostasis pathways were substantially reduced or no longer detected within the BHB-responsive interactome. Network enrichment analysis of the transporter perturbation datasets instead highlighted broader functional modules, including RNA metabolism and protein assembly pathways, suggesting a shift toward general cellular stress responses.

In contrast, inhibition of the glial transporter chk produced a more limited effect on the C99-associated protein network. Vesicle trafficking and proteostasis-related modules remained prominently associated with the C99 interactome, consistent with the functional pathways identified in the initial interactome analysis (Fig. 4c).

Together, these findings suggest that neuronal ketone uptake is required for BHB-dependent remodelling of the C99-associated protein network.

## VPS35 mediates the rescue of vesicle trafficking and autophagy defects

Proteomic network analysis identified VPS35, a core component of the retromer complex involved in endosomal trafficking, as a candidate hub protein within the C99-associated interactome. To determine whether VPS35 contributes to the protective effects of ketone treatment, we examined autophagy and mitochondrial quality control following RNAi-mediated depletion of VPS35 in the C99 model [33].

In control animals, BHB treatment partially restored autophagy reporter signals in C99-expressing retinas. However, RNAi-mediated knockdown of VPS35 abolished these effects. Specifically, VPS35 depletion prevented the reduction of LC3/Atg8a or Ref(2)P accumulation typically observed following BHB treatment, indicating persistent accumulation of autophagic cargo (Fig. 4a–d). Similarly, the improvement in mitochondrial turnover detected using the mito-QC reporter was lost upon VPS35 depletion (Fig. 4e,f). These findings indicate that VPS35 contributes to the BHB-dependent restoration of autophagy and mitochondrial quality control pathways in the C99 model.

Given the prominent vesicular abnormalities observed in C99-expressing photoreceptors, we next examined whether VPS35 influences vesicle morphology at the ultrastructural level. Transmission electron microscopy revealed extensive accumulation of enlarged vesicular compartments in C99 retinas. While BHB treatment normally reduced vesicular burden in this model, VPS35 depletion prevented this normalization (Fig. 4h–k, Red box). Instead, photoreceptors lacking VPS35 retained enlarged vesicular structures resembling those observed in untreated C99 tissue. Quantitative analysis confirmed that vesicular area remained elevated following VPS35 knockdown.

# Discussion

Alzheimer's disease (AD) has traditionally been defined by the accumulation of extracellular amyloid-β plaques and intracellular tau pathology[3,34]. However, increasing evidence indicates that intracellular organelle dysfunction precedes these classical hallmarks, suggesting that early disturbances in membrane trafficking and protein clearance may represent initiating events in disease pathogenesis[13,35]. Among these alterations, enlargement of endosomal compartments, accumulation of autophagic vacuoles, and lysosomal dysfunction are among the earliest cellular abnormalities detected in AD brains[6]. Our findings support this emerging view by demonstrating that expression of the amyloid precursor protein fragment C99 in Drosophila induces extensive vesicular accumulation, impaired autophagic turnover, and disruption of mitochondrial quality control.

Recent studies suggest that APP C-terminal fragments such as C99 may play a more direct role in cellular toxicity than previously appreciated. Accumulation of APP-CTFs has been shown to destabilize endolysosomal homeostasis and interfere with membrane trafficking pathways independently of extracellular amyloid deposition[6,36,37]. In agreement with this model, we observed pronounced vesicular abnormalities and impaired mitochondrial turnover in C99-expressing photoreceptors, indicating that C99 accumulation alone is sufficient to disrupt intracellular trafficking and organelle quality control. These results support a growing body of evidence suggesting that C99-driven organelle stress may represent an early pathogenic mechanism in AD.

Our ultrastructural analyses revealed that C99 expression leads to the accumulation of enlarged vesicular structures containing dense luminal material, consistent with defects in degradative vesicle maturation[38,39]. Such vesicular overload is a hallmark of impaired autophagy–lysosomal function in neurons, which rely heavily on efficient organelle clearance to maintain cellular homeostasis[40]. Together with the observed defects in mitochondrial turnover, these findings suggest that C99 accumulation disrupts multiple interconnected pathways involved in vesicle trafficking, autophagy, and mitochondrial quality control.

A central observation of this study is that ketone treatment restores autophagic and mitochondrial turnover in the C99 model. Ketone bodies such as β-hydroxybutyrate (BHB) have recently attracted attention as neuroprotective metabolites capable of modulating mitochondrial metabolism, redox balance, and cellular stress responses[41–43]. In our model, BHB treatment reduced autophagic cargo accumulation, restored mitochondrial turnover, and normalized vesicular ultrastructure. Importantly, these effects required neuronal ketone transport, indicating that the protective mechanism is mediated by neuron-intrinsic metabolic signaling rather than systemic metabolic changes. These results provide a mechanistic framework for understanding how ketogenic interventions may influence neuronal homeostasis in neurodegenerative disease.

Proteomic analysis further revealed that ketone treatment is associated with remodeling of the C99-associated interactome, highlighting a network of proteins involved in vesicle trafficking and proteostasis. Intracellular trafficking of APP and its processing enzymes has long been recognized as a critical determinant of amyloidogenic processing and cellular toxicity [44]. The identification of trafficking-related modules within the C99 interactome therefore reinforces the idea that disruption of membrane trafficking pathways is a key driver of pathology in this model.

Among the candidate regulators identified in our network analysis, VPS35 emerged as a central hub linking vesicle trafficking and autophagy pathways. VPS35 is a core component of the retromer complex, which mediates retrograde sorting of membrane proteins from endosomes to the trans-Golgi network and thereby maintains endosomal homeostasis[32,45,46]. Retromer dysfunction has been implicated in several neurodegenerative diseases, including Alzheimer's and Parkinson's disease, where it contributes to defects in endosomal recycling, mitochondrial dynamics, and autophagy pathways [45]. Consistent with these functions, RNAi-mediated depletion of VPS35 in our model abolished the beneficial effects of ketone treatment on autophagy reporters, mitochondrial turnover, and vesicle morphology. These findings suggest that VPS35-dependent trafficking pathways are required for the ketone-mediated restoration of vesicle homeostasis.

Several limitations of the present study should be considered. Although VPS35 depletion suppressed the protective effects of ketone treatment, RNAi approaches cannot fully establish genetic necessity, and further studies using complementary genetic strategies will be required to define the precise molecular mechanisms connecting ketone signaling to retromer function. Additionally, while the Drosophila model provides a powerful system for dissecting early cellular mechanisms of C99 toxicity, the extent to which similar pathways operate in mammalian neurons remains to be determined.

In summary, our study identifies a metabolic–trafficking axis linking ketone signaling, vesicle trafficking, and autophagy regulation in a C99-driven model of neurodegeneration. These findings support an emerging view in which intracellular trafficking defects represent early pathogenic events in Alzheimer's disease and suggest that metabolic interventions capable of restoring retromer-dependent trafficking pathways may represent promising therapeutic strategies.

# Data availability

The raw data supporting the conclusions of this article will be made available by the authors on request.

# Acknowledgements

The authors acknowledge the instruments and expertise of Microscopy Australia (ROR: 042mm0k03) at Adelaide Microscopy, University of Adelaide, enabled by NCRIS, university, and state government support.

The authors acknowledge the instruments and expertise of Adelaide Proteomics Centre, with special thanks to Associate Professor Tara Pukala for her invaluable guidance on mass spectrometry.

This research was made possible through the invaluable support of the Adelaide *Drosophila* community, the *Drosophila* Facility at the University of Adelaide, and its outstanding researchers, Professor Robert Richards and Dr. Louise O'Keefe. We also extend our heartfelt condolences to memory of Dr. Louise O'Keefe.

This manuscript was edited using Overleaf, incorporating suggestions from language models, including Writefull's model and GPT-based models. AI-assisted tools were utilized to improve grammar, clarity, and academic style. All modifications were reviewed and accepted by the authors to ensure accuracy and adherence to the intended scientific meaning.

# Funding

No funding was received towards this work.

# Competing interests

The authors report no competing interests.

# Clinical trial number

Not applicable

## Authors' contributions

HH conducted the main body of the study, including *Drosophila* data collection, processing, and code development. KX was responsible for daily fly maintenance and carried out all blinded procedures during data acquisition. ML supervised HH and KX, providing essential equipment and technical guidance throughout the research. All authors read and approved the final manuscript.

# Figure legends

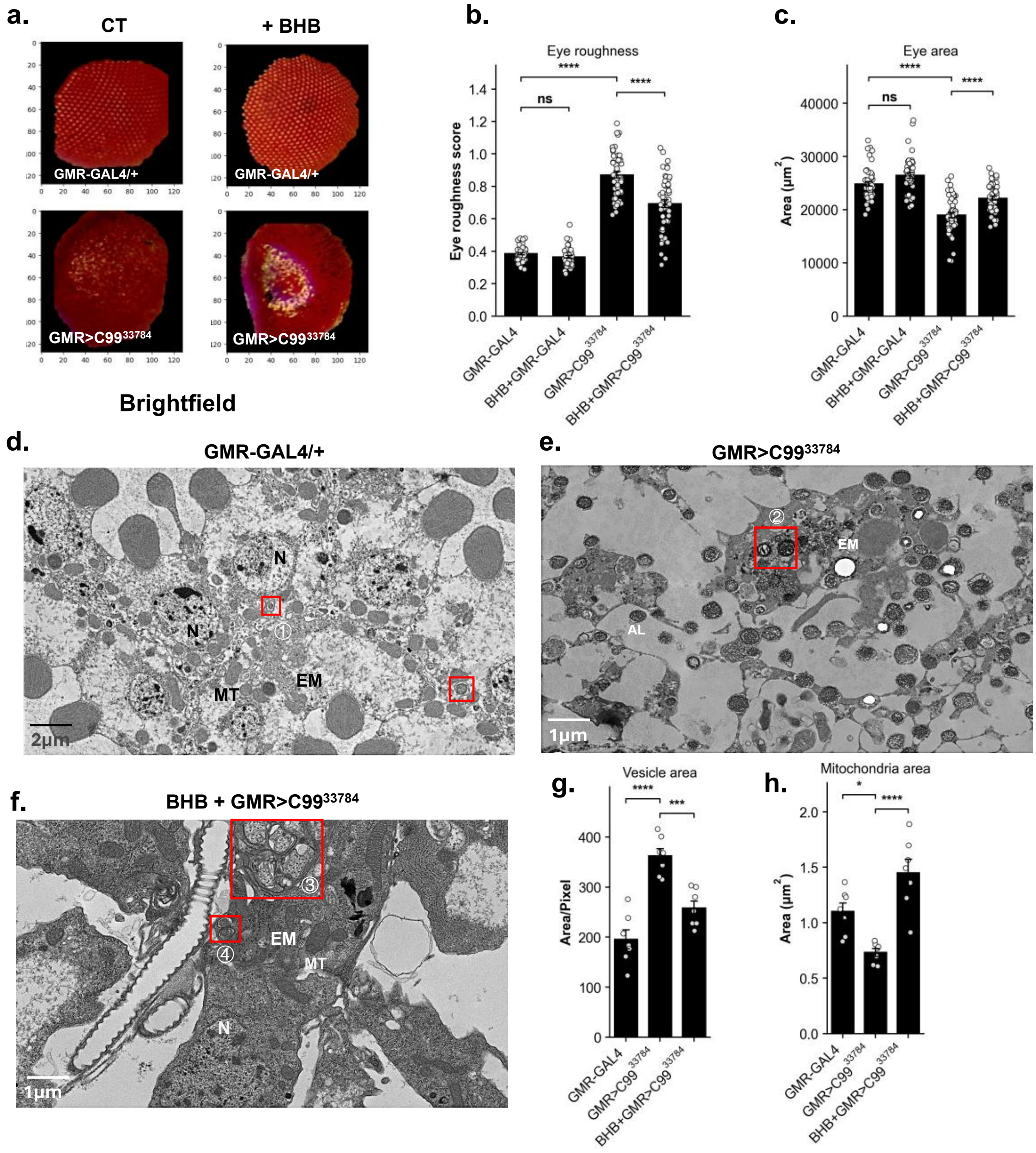

**Figure 1 C99 expression induces retinal degeneration and vesicular accumulation in Drosophila photoreceptors.** (a) Representative images of adult *Drosophila* compound eyes. Control flies (GMR-GAL4/+) display a highly ordered ommatidial lattice, whereas expression of human C99 (GMR>C99) results in a pronounced rough eye phenotype indicative of retinal degeneration. Eye surface topology was visualized using computational surface mapping. (b) Quantification of eye roughness scores using a convolutional neural network–based phenotyping pipeline. C99 expression significantly increased eye roughness compared with controls, while β-hydroxybutyrate (BHB) treatment partially suppressed this phenotype. (c) Quantification of eye area. C99 expression reduced eye size relative to controls, and this reduction was partially restored by BHB treatment. (d–f) Transmission electron microscopy (TEM) analysis of photoreceptor cells in the *Drosophila* retina. Control photoreceptors exhibit normal cellular organization with clearly identifiable nuclei (N), mitochondria (MT), and endomembrane structures (EM). In contrast, C99-expressing photoreceptors display extensive accumulation of enlarged vesicular compartments and abnormal membrane-bound structures. Red boxes highlight double-membrane vesicles morphologically resembling autophagic or autophagosome-like structures. (g) Quantification of vesicular area within photoreceptor cells, demonstrating a significant increase in vesicular accumulation in C99-expressing retinas, which was partially reduced by BHB treatment. (h) Quantification of mitochondrial area, revealing altered mitochondrial morphology in C99-expressing photoreceptors. Data are presented as mean ± s.e.m., with each point representing an individual biological replicate. Statistical significance was determined using one-way ANOVA followed by Tukey's post hoc test. ns, not significant; *$p < 0.05$; **$p < 0.01$; ***$p < 0.001$; ****$p < 0.0001$. $n_{a-c}$=49-50, $n_{d-f}$=10-12.

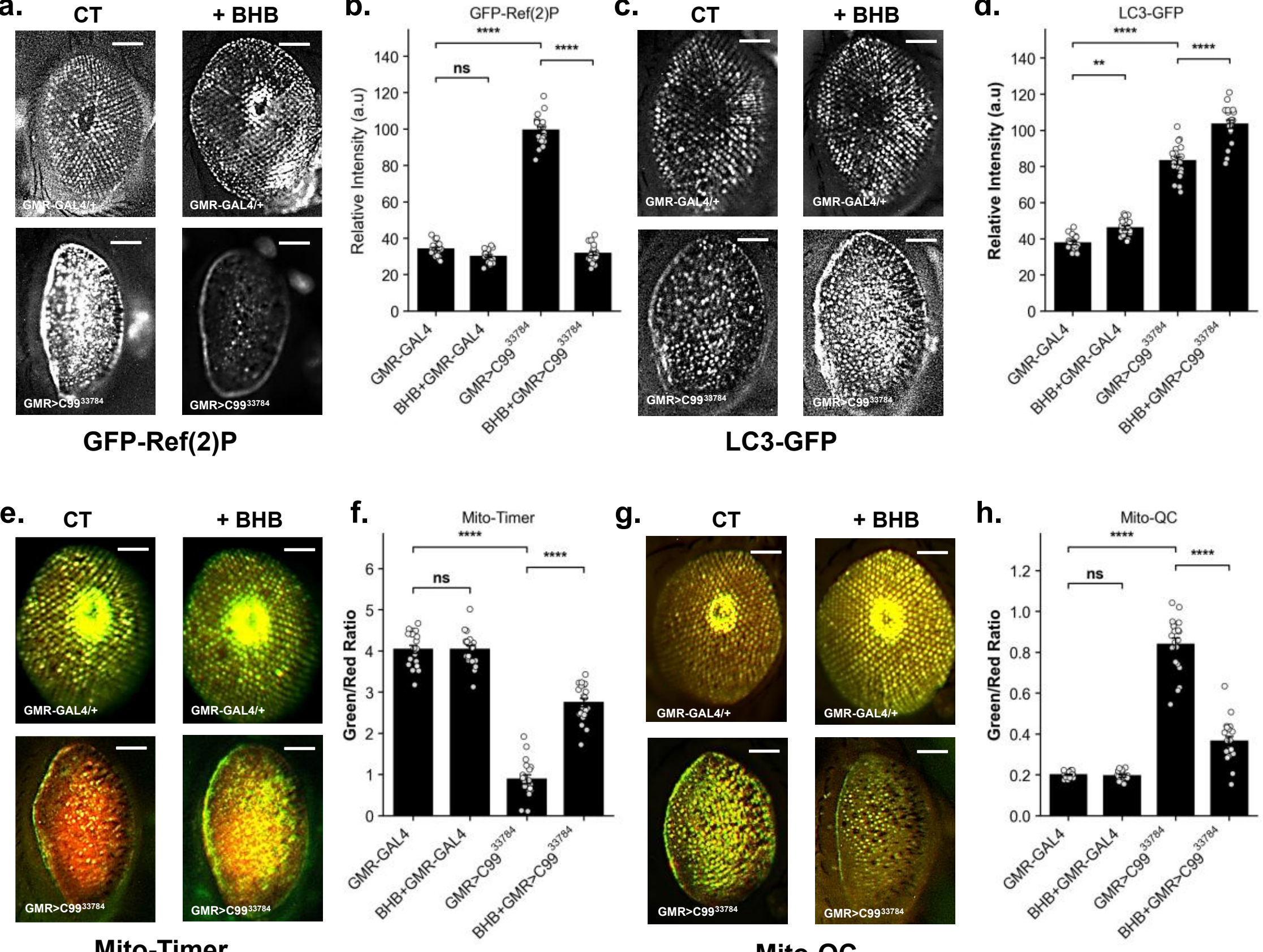

**Figure 2 Ultrastructural and functional assessment of vesicles, mitochondria, and autophagy in *Drosophila* eye models. (a–b)** GFP-Ref(2)P reporter analysis in adult *Drosophila* eyes. Representative fluorescence images and quantification of GFP-Ref(2)P intensity are shown. Expression of human C99 (GMR>C99) resulted in a marked accumulation of Ref(2)P compared with controls, indicating impaired clearance of autophagic cargo. Treatment with β-hydroxybutyrate (BHB) significantly reduced Ref(2)P accumulation. **(c–d)** LC3-GFP (Atg8a) reporter analysis. Representative images and quantification of LC3-GFP signal intensity in photoreceptors are shown. C99 expression increased LC3-positive structures relative to controls, consistent with accumulation of autophagosome-associated membranes. BHB treatment did not reduce LC3 reporter levels, suggesting that ketone treatment does not simply suppress autophagosome formation. **(e–f)** MitoTimer analysis of mitochondrial age. Representative images and quantification of the green/red fluorescence ratio are shown. C99 expression markedly decreased the MitoTimer ratio, consistent with accumulation of aged mitochondria. BHB treatment partially restored the ratio. **(g–h)** mito-QC reporter analysis of mitophagy. Representative images and quantification of the green/red fluorescence ratio are shown. C99 expression significantly increased the mito-QC ratio, consistent with reduced mitochondrial delivery to lysosomes. BHB treatment partially reversed this phenotype. Data are presented as mean ± s.e.m., with each point representing an individual biological replicate. Statistical significance was determined using **one-way ANOVA followed by Tukey's post hoc test**.ns, not significant; *$p < 0.05$; **$p < 0.01$; ***$p < 0.001$; ****$p < 0.0001$. Scale bars: 100 μm. n=20.

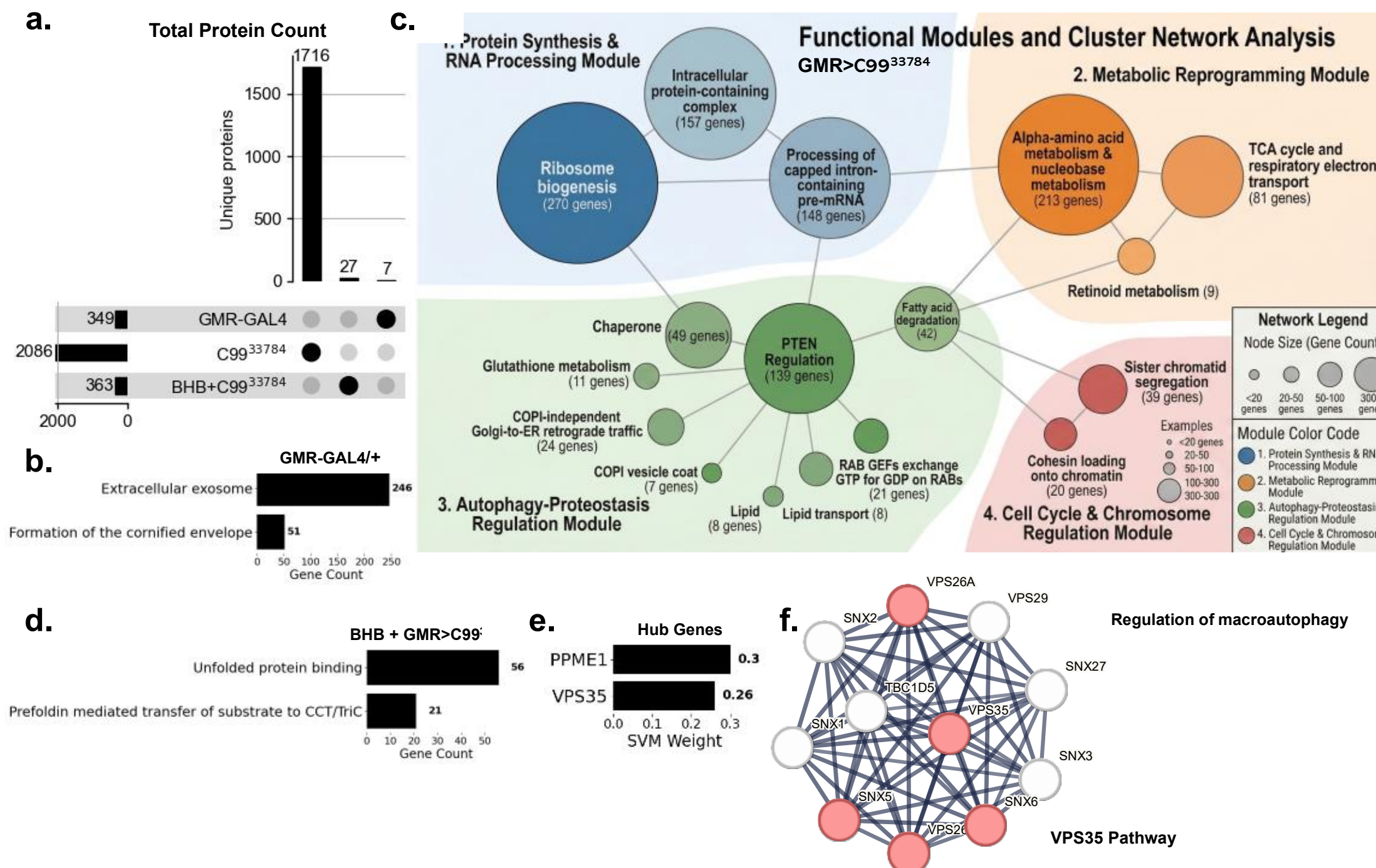

**Figure 3 Proteomic and network analyses identify candidate regulatory nodes within the C99-associated interactome.** (a) Identification of proteins associated with C99-containing complexes. Co-immunoprecipitation coupled with mass spectrometry (Co-IP/MS) was performed using flies expressing C99 in photoreceptors. After subtraction of proteins detected in control samples (GMR-GAL4/+), a set of C99-associated proteins was identified. The bar graph shows the number of proteins detected in control, C99, and BHB-treated C99 samples. (b) Functional enrichment analysis of the C99-associated interactome. Gene ontology analysis revealed enrichment of proteins involved in extracellular vesicles and membrane-associated compartments. (c) Network-based functional module analysis of proteins identified in the C99 interactome. Enriched pathways were organized into four major functional modules, including protein synthesis and RNA processing, metabolic reprogramming, autophagy–proteostasis regulation, and cell structure and chromosome regulation. Node size represents gene count within each pathway. (d) Functional enrichment analysis of proteins associated with the C99 interactome under BHB treatment. Enriched pathways included unfolded protein binding and chaperone-mediated protein folding pathways. (e) Machine learning–based prioritization of candidate hub genes. Human Alzheimer's disease transcriptomic data (GSE5281) were analyzed using an integrated machine learning framework combining random forest feature ranking, autoencoder-based feature extraction, and support vector machine classification. Gene importance scores were used to prioritize candidate regulatory nodes among proteins identified in the C99 interactome. (f) Top candidate hub genes identified through integrative analysis. VPS35 and PPME1 emerged as high-scoring candidates based on SVM-derived importance weights. (g) Network representation of candidate hub proteins within the C99-associated protein interaction network. VPS35 is centrally positioned within a cluster containing multiple retromer and sorting nexin components, suggesting a potential role in vesicle trafficking and proteostasis pathways.

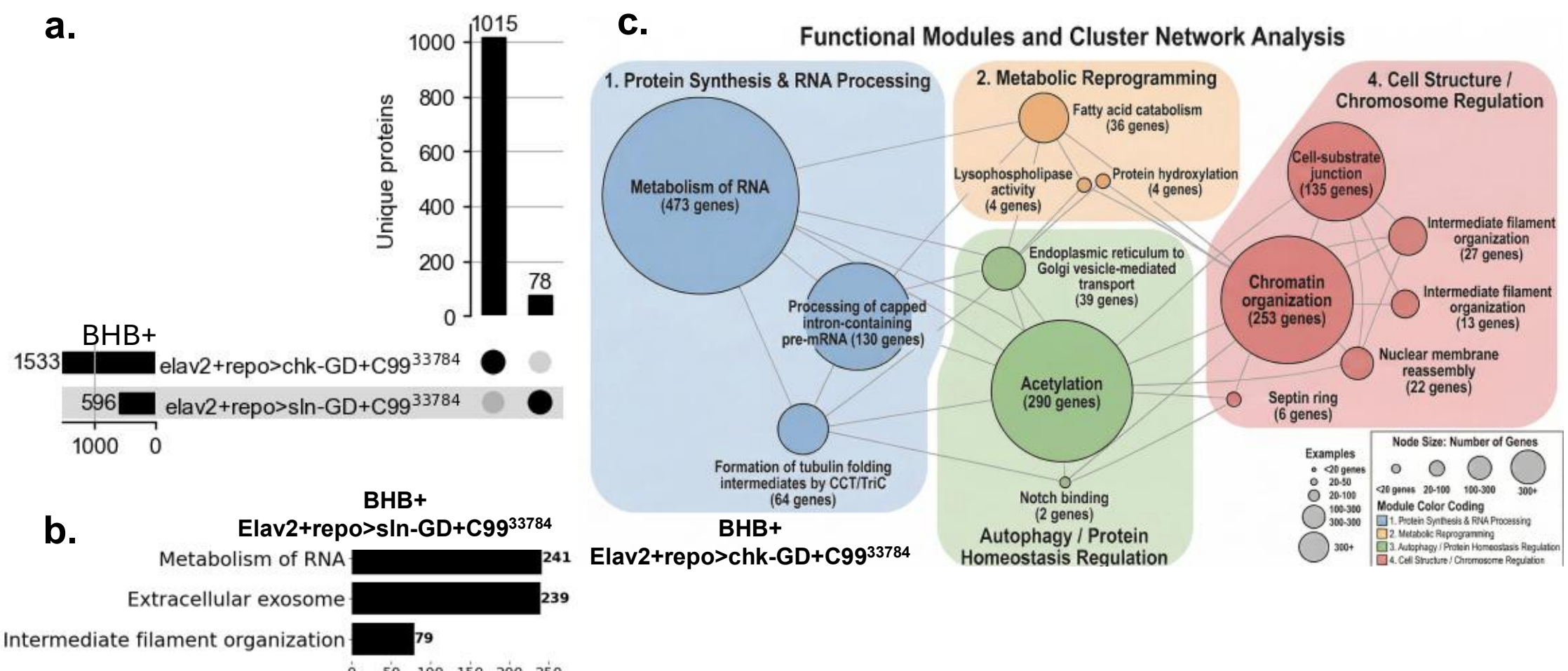

**Figure 4 Neuronal uptake of ketone bodies shapes the C99-associated interactome.** (a) Comparison of proteins detected in the C99-associated interactome under BHB treatment following knockdown of ketone body transporters. RNAi targeting the neuronal ketone transporter sln or the glial transporter chk was expressed using the elav2+repo driver in C99-expressing flies. The number of proteins identified under each condition is shown. (b) Functional module and cluster network analysis of proteins identified in the interactome following neuronal ketone transporter knockdown. Enriched pathways were organized into four major functional modules, including protein synthesis and RNA processing, metabolic reprogramming, autophagy–protein homeostasis regulation, and cell structure and chromosome regulation. Node size represents gene count within each pathway. (c) Gene ontology enrichment analysis of interactome changes associated with transporter knockdown. In flies with neuronal sln knockdown, enriched pathways included RNA metabolism and extracellular vesicle–related processes. In contrast, knockdown of the glial transporter chk preferentially enriched pathways associated with intermediate filament organization.

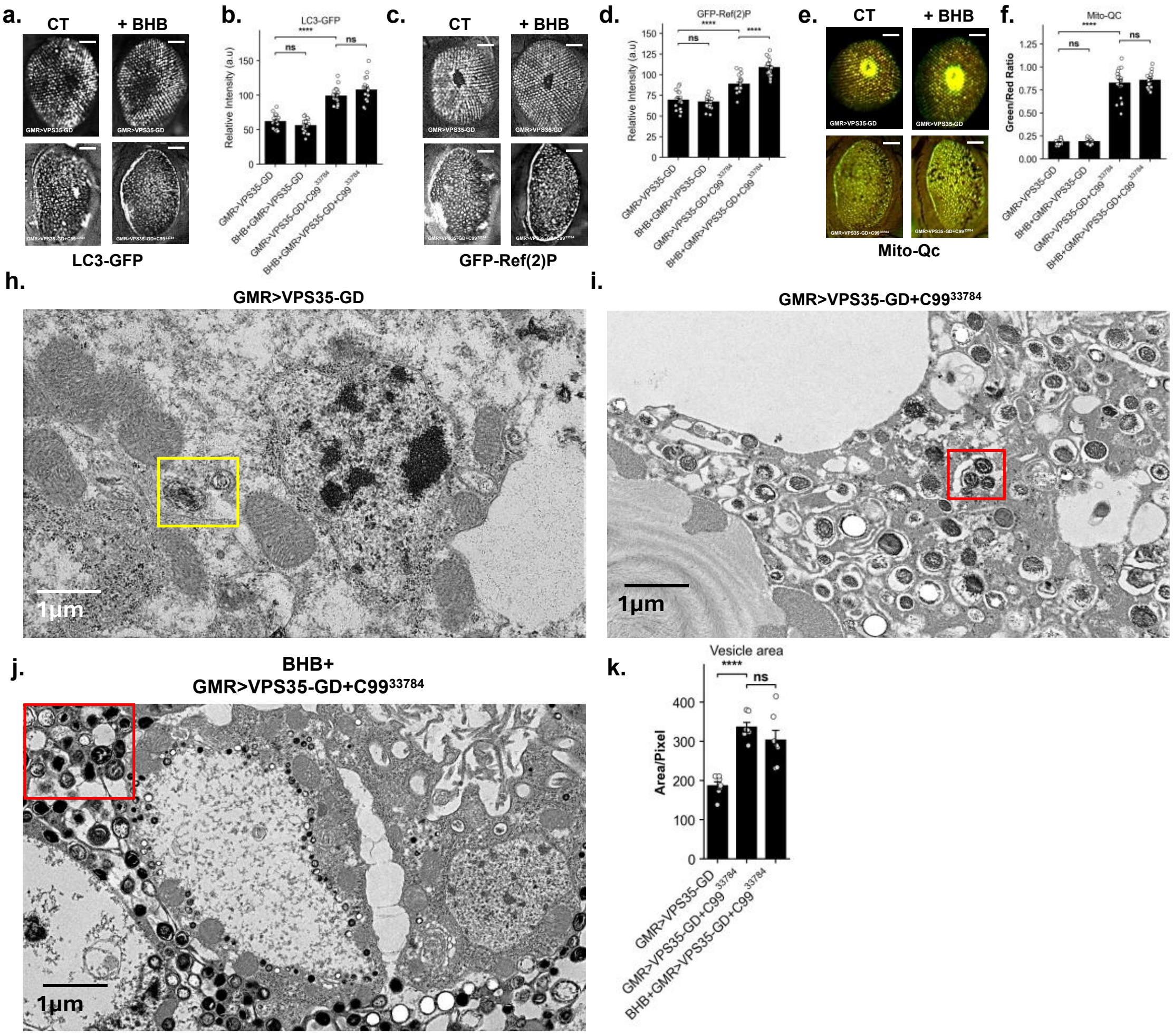

**Figure 5 VPS35 depletion abolishes the protective effects of ketone treatment on autophagy and vesicular homeostasis.** (a–b) LC3-GFP (Atg8a) reporter analysis in adult *Drosophila* eyes. Representative fluorescence images and quantification of LC3-GFP intensity are shown. Knockdown of VPS35 resulted in increased LC3-positive structures in C99-expressing retinas. BHB treatment did not reduce LC3 reporter levels under VPS35 knockdown conditions. (c–d) GFP-Ref(2)P reporter analysis. Representative images and quantification of GFP-Ref(2)P intensity are shown. VPS35 knockdown resulted in accumulation of Ref(2)P in C99-expressing retinas, indicating impaired autophagic cargo clearance. BHB treatment failed to reduce Ref(2)P levels when VPS35 was depleted. (e–f) mito-QC reporter analysis of mitophagy. Representative images and quantification of the green/red fluorescence ratio are shown. VPS35 depletion significantly impaired mitochondrial delivery to lysosomes in C99-expressing photoreceptors. BHB treatment did not restore mito-QC reporter activity under VPS35 knockdown conditions. (g–i) Transmission electron microscopy (TEM) analysis of photoreceptor cells. VPS35 knockdown in C99-expressing retinas resulted in extensive accumulation of vesicular compartments and abnormal membrane-bound structures. The yellow boxes represent typical examples of two-membrane autophagy that can be referenced. Red boxes highlight enlarged vesicular structures containing multilamellar or electron-dense material. (j) Quantification of vesicular area within photoreceptor cells, demonstrating persistent vesicular accumulation following VPS35 depletion. BHB treatment did not significantly reduce vesicle area under VPS35 knockdown conditions. Data are presented as mean ± s.e.m., with each point representing an individual biological replicate. Statistical significance was determined using one-way ANOVA followed by Tukey's post hoc test. ns, not significant; *$p < 0.05$; **$p < 0.01$; ***$p < 0.001$; ****$p < 0.0001$. $n_{a\text{-}f}=20$,$n_{g\text{-}i}=10$.

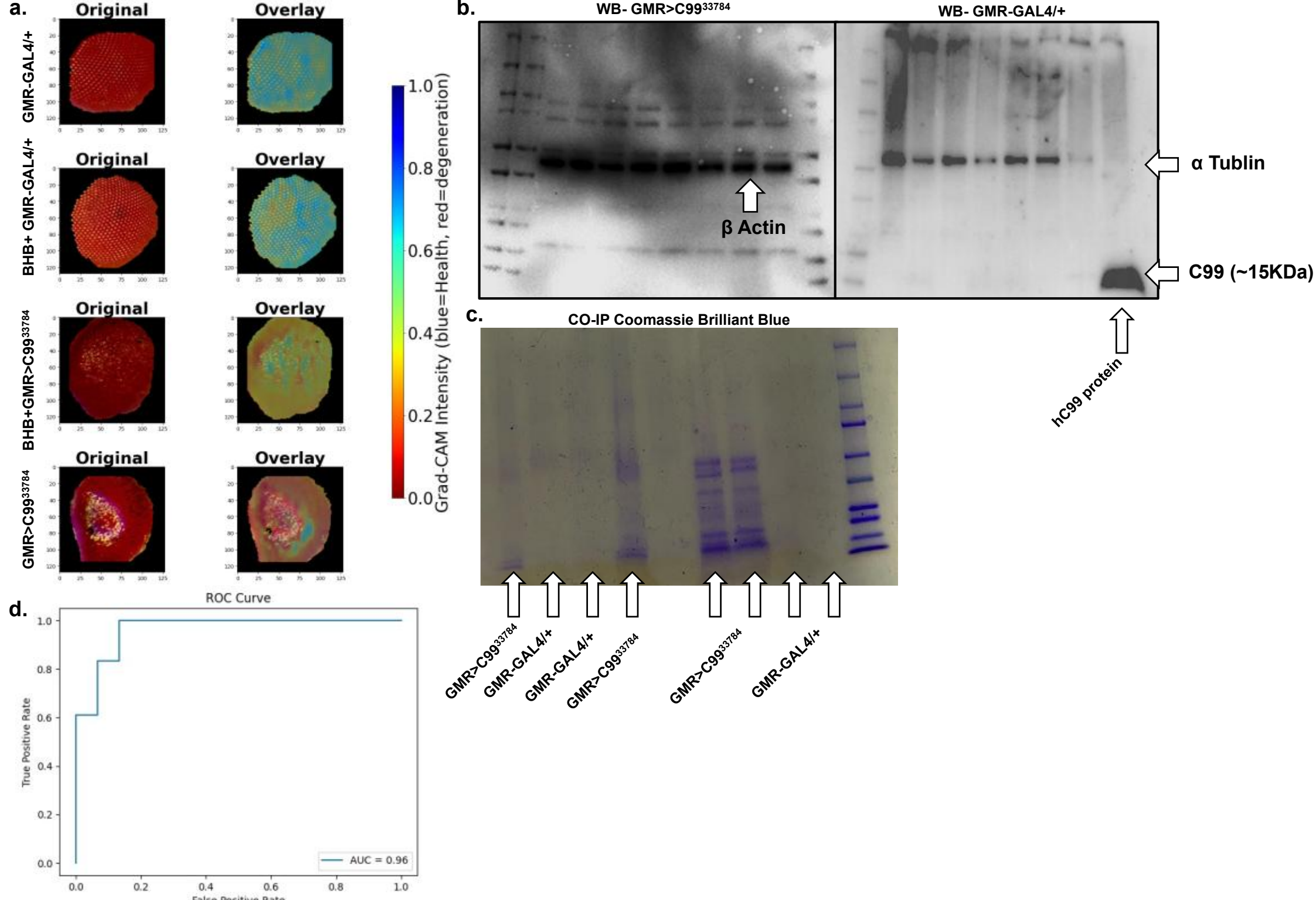

**Figure S1 Validation of image analysis and proteomic workflows. (a)** Grad-CAM visualization of the convolutional neural network used for automated eye roughness scoring. Representative examples of original compound eye images and Grad-CAM overlays are shown. The heatmap indicates regions contributing most strongly to the classification model (blue, healthy regions; red, degenerated regions). **(b)** Validation of C99 expression and immunoprecipitation efficiency. Western blot analysis showing C99 expression in experimental samples. β-actin was used as a loading control. **(c)** Coomassie Brilliant Blue staining of SDS–PAGE gels showing proteins isolated by co-immunoprecipitation prior to mass spectrometry analysis. Arrows indicate lanes corresponding to samples used for downstream proteomic identification. **(d)** Receiver operating characteristic (ROC) curve showing the performance of the machine learning framework used for hub gene prioritization. The model was trained on human Alzheimer's disease transcriptomic data (GSE5281) using an integrated pipeline combining random forest feature selection, autoencoder-based feature extraction, and support vector machine classification. The area under the curve (AUC) indicates strong classification performance.